\documentclass[useAMS,usenatbib]{mn2e}

\usepackage{graphicx}
\usepackage{amssymb}
\usepackage{subfigure}

\def\RP{{r_{\rm c}}}
\def\rs{{r_{\rm s}}}

\title{A Tale Twice Told: The Luminosity Profiles of the Sagittarius Tails}
\author[M. Niederste-Ostholt et al.]
{M.Niederste-Ostholt$^{1}$\thanks{E-mail:mno, vasily, nwe@ast.cam.ac.uk},
  V. Belokurov$^{1}$, N.W. Evans$^{1}$\\
  $^{1}$Institute of Astronomy, Madingley Rd, Cambridge, CB3 0HA\\}

\begin{document}

\date{May 2011}

\pagerange{\pageref{firstpage}--\pageref{lastpage}} \pubyear{2009}

\maketitle

\label{firstpage}

\begin{abstract}
  The Sagittarius dwarf galaxy is the archetype of a tidally
  disrupting system. Both leading and trailing tails can be observed
  across at least 180 degrees of the sky and measurements of their
  luminosity density profiles have recently become available.  Using
  numerical simulations, we explore the factors that control the
  appearance of such profiles.
 
  We use two possible models for the Sgr progenitor. The first is a
  one-component Plummer model, which may represent either a dark
  matter free progenitor, or one in which pre-existing dark matter has
  already been largely stripped. The second is a two-component model
  in which the stars are represented by a Hernquist sphere embedded in
  a cosmologically modish Navarro-Frenk-White dark halo. Disruption of
  the models in the Milky Way galaxy provides us with two tellings of
  the tale of the formation of the Sgr stream. The initial
  disintegration of the baryons proceeds more slowly for the
  two-component models because of the protective cocoon of dark
  matter. Once this has been stripped, though, matters proceed
  apace. In both cases, the profiles after $\sim 6$ pericentric
  passages provide good matches to the observational data, but the
  tails are more extended for the two-component models.

  The leading and trailing tails are symmetric at apocentre or
  pericentre. At other orbital phases, asymmetries are present, as
  tails are compressed as they approach apocentre and stretched out as
  they approach pericentre.  There may exist density enhancements
  corresponding to such pile-ups which may be observable in current
  survey data. We re-visit the calculation of Niederste-Ostholt et
  al. (2010) and slightly revise upwards the luminosity of the Sgr
  progenitor to $9.9-14.4\times10^7 L_{\odot}$ based on insights from
  the simulations.
\end{abstract}

\begin{keywords}
  galaxies: dwarf galaxies -- galaxies: individual (Sagittarius) --
  galaxies: simulations
\end{keywords}

\section{Introduction}

This paper provides two tellings of the tale of the Sagittarius
(Sgr) galaxy.

Ever since its discovery~\citep{Ibata:1995p334}, the Sgr dwarf has
been recognised as a touchstone of Galactic astronomy. It provides an
ongoing example of the merging and accretion of galaxies, one of the
fundamental processes through which structure is built in the
Universe.  It provides one of the best measures of the Galactic
potential and mass, as its tidal debris can be traced to distances of
possibly $\sim 90$ kpc \citep{Newberg:2003p338, Belokurov:2006p466,
  2011arXiv1111.7042K}.  The disruption of Sgr and the spilling of its
stars and globular clusters into the Galaxy provides a sizeable
fraction, perhaps as much as a quarter, of the entire stellar halo of
the Milky Way by luminosity (using numbers from \citet{Bell:2008p2889} or
\citet{2011MNRAS.416.2903D} and \citet{NiedersteOstholt:2010p2890}).

Given its fundamental role, the disruption of the Sgr dwarf and the
formation of its tidal tails has been a popular topic for detailed
numerical investigations. Motivated by the proximity of the remnant to
the Galactic centre, simulations as to the effect of Milky Way tides
on Sgr began soon after its discovery \citep{Johnston:1995p2124}.  The
aims of subsequent simulations are largely threefold. First, some
investigators concentrated on understanding the Sgr progenitor
\citep[e.g.][]{Velazquez:1995p2123,Ibata:1998p2826,Lokas:2010p3684}.
The work of \citet{Jiang:2000p523} nicely summarizes the degeneracy in
the properties of the progenitor, with the then available data being
compatible with a very broad range of masses and orbital histories.
Secondly, there is a body of work that has studied what can be
inferred about the structure of the Milky Way, particularly its dark
halo. This though has not led to clear-cut conclusions on the shape of
the dark halo which has been variously claimed as spherical, oblate,
prolate or triaxial~\citep[e.g.][]
{Helmi:2004p524,Helmi:2004p525,Johnston:2005p3031,Fellhauer:2006p340,Law:2010p2807}.
Thirdly, some investigators have been concerned primarily with the
properties of the debris itself, particularly as positions, kinematics
and metallicities of presumed stream members have gradually become
available from wide area surveys like 2MASS and SDSS
\citep[e.g.][]{Johnston:1998p2079,MartinezDelgado:2004p1138,Law:2005p522,Law:2010p2807}.

Recently, \citet{NiedersteOstholt:2010p2890} have used the extant
2MASS and SDSS survey data to produce luminosity density profiles of
both the leading and trailing tails. By extrapolating the observed
profiles both near the core and along the orbit, the luminosity of the
undisrupted system can be determined. These profiles have not yet been
explored by simulations. In this contribution, we provide with the aid
of N-body models two re-tellings of the tale of the Sgr galaxy.  We
are interested in constraining the dark matter content of the
progenitor, as well as understanding what controls the appearance of
the tail luminosity profiles.

Section 2 is primarily technical and gives details of the set-up of
the simulations together with the analyses performed on the
output. Section 3 provides the first telling of the tale with
simulations in which the dark matter shadows the light in the
progenitor. Section 4 provides the second telling of the tale with
dark matter dominated progenitors. Finally, we draw our conclusions in
Section 5.

\begin{figure*}
	\centering
	\includegraphics[width=1 \textwidth]{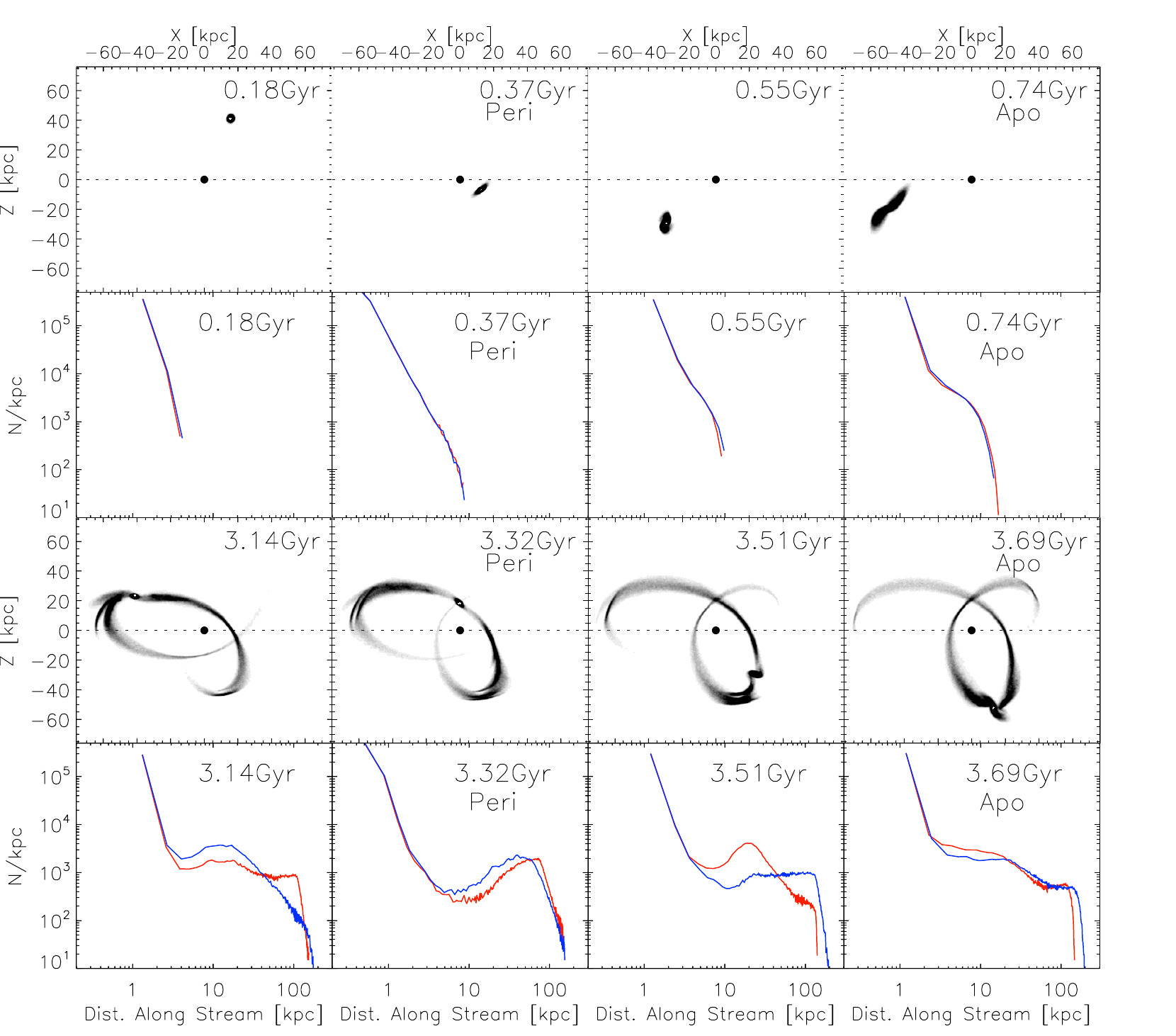}
        \caption[Sample Simulation Profiles]{Snapshots and profiles of
          a typical simulation looking face-on into the orbital
          plane. The satellite has a Plummer radius 350 pc and mass
          $10^8 M_{\odot}$.  The plots run forward in time from left
          to right and top to bottom.  The dot shows the Galactic
          centre, which is centre of the Cartesian axes ($X,Z$) with
          $Z=0$ defining the Galactic plane. The profiles represent
          the number density of particles along the stream with the
          leading arm in red and the trailing arm in blue.  The early
          evolution of the satellite and its accompanying profiles are
          shown in the top two rows. The leading and trailing tails
          begin forming symmetrically and the profiles lie on top of
          each other. The bottom two rows show the later evolution of
          the satellite. Asymmetries begin to occur as the tails fill
          appreciable portions of the orbital path, between apocentre
          and pericentre.  Overdensities occur as stars slow down near
          apocentre and the tail is compressed, whereas lower density
          areas occur the the tail is approaching pericentre and is
          stretched out. The profiles remain symmetric when Sgr is
          observed at either apocentre or pericentre. Asymmetries
          between the leading and trailing arms appear when Sgr is at
          intermediate positions along its orbit.}
\label{fig:snaps}
\end{figure*}

\section{Set-Up and Assumptions}

An N-body satellite is placed on an orbit in a static Milky Way-like
potential consisting of a \citet{Miyamoto1975PASJ...27..533M} disc
($M_{\rm d}=7.5\times10^{10} M_{\odot}$, radial scale length $=3.5$
kpc, vertical scale length$=0.3$ kpc), a \citet{Hernquist:1990p1309}
bulge ($M_{\rm b}=1.3\times10^{10} M_{\odot}$, scale radius $=1.2$
kpc), and a spherically symmetric logarithmic halo (circular velocity
$186$ kms$^{-1}$, scale size $12$ kpc). Similar potentials have been
used before ~\citep[see
e.g.][]{Johnston:1999p1533,Law:2005p522,Fellhauer:2006p340}.

We explore two possibilities for the progenitor Sgr. In the first
telling of the tale, the satellite is represented by a Plummer sphere
of characteristic lengthscale $\RP$ composed of $10^6$ particles. This
single component approach follows many previous papers, including
\citet{Law:2010p2807} who are able to reproduce a wide array of
observations. This mass-follows-light model can be interpreted as
representing either a situation in which all dark matter has been
stripped before the baryons begin significant disruption, or in which
the satellite had little dark matter to begin with.  In the second
telling, the Sgr galaxy is embedded in a dark matter
halo. Specifically, we use a two-component model consisting of baryons
distributed in a Hernquist sphere of lengthscale $\RP$ embedded in an
Navarro-Frenk-White (NFW) dark matter halo with lengthscale $\rs$
(c.f. Fellhauer et al. 2007). In both tellings, the models are
released in the same Milky Way potential with the same initial
conditions. They have the same characteristic lengthscale and mass for
the baryonic component. Their initial mean velocity dispersions are
different (namely $\sim 15$ kms$^{-1}$ and $\sim 19$ kms$^{-1}$), but
they both have mean dispersions consistent with the observations \citep{Ibata:1997p532} at
the timestep corresponding to the present-day epoch ($\sim 10$
kms$^{-1}$ and $\sim 11$ kms$^{-1}$ respectively).

The size and mass of Sgr prior to disruption can only be estimated.
However, based on the luminosity determined in
\citet{NiedersteOstholt:2010p2890}, a mass of at least $10^8$
$M_{\odot}$ for the baryonic component is justifiable.  Although we
will consider the effects of the satellite mass and size on the shape
of the luminosity profiles, our benchmark one and two-component models
have a baryonic mass of $6.4\times10^8 M_{\odot}$ and a characteristic
radius $\RP = 850$ pc~\citep[c.f.][]{Law:2010p2807}.

The evolution of the satellite as it tidally disrupts is followed by
looking at snap-shots of the particle distribution at various
locations along the orbit.  The disruption is simulated with the
particle-mesh code {\tt SuperBox} \citep{Fellhauer:2000p539}. For
one-component models, it uses three nested grids of $64^3$ cells each
centred on the highest density region of the satellite galaxy. The
inner high-resolution grid covers the central region of the satellite
with the grid spanning $2\times \RP$. The medium resolution grid
covers the entire satellite ($10\times \RP$). The low resolution grid
covers $65\times R_{\rm cutoff}$, where $R_{\rm cutoff}=5 \RP$ is the
artificial cut-off radius imposed on the initial distribution of the
dwarf galaxy. The simulation steps forward with a constant time-step
size of $\Delta t= 1.3$ Myr, which leads to a total energy
conservation better than $1\%$ when the satellite is evolved in
isolation for $\sim 10$ Gyr. For the two-component models, there are
two sets of grids -- the first covering the baryons and the second the
dark matter particles. The sizes are set up relative to the baryonic
characteristic scale $\RP$ and the NFW scale radius $\rs$ in an
analogous way to the one-component models.

The orbit has initial conditions $(x,y,z)=(0,0,55)$ kpc and
$(u,v,w)=(90,0,0)$ kms$^{-1}$ which result in a pericentre of
$\approx15$ kpc and an apocentre $\approx55$ kpc in line with
observations of Sgr \citep[e.g.][]{Ibata:1998p2826,
  Johnston:1999p1658,Ibata:2001p296,Law:2005p522}.  Radial and
transverse velocity measurements of the main body of Sgr, as well as
2MASS observations of the trailing arm, indicate that the Sgr dwarf is
currently located close to pericentre. The orbital period is
approximately $0.7$ Gyr. During each period, eight snapshots are taken
in equally spaced time-steps, chosen so that one snap-shot is at
apocentre and one at pericentre.  The orbit lies in the plane
perpendicular to the galactic disk, which makes the subsequent
analysis considerably more straightforward. Observations of Sgr debris
support the notion that the material torn from the satellite is
confined to a plane \citep{Majewski:2003p337}, with an orbital pole at
$(\ell,b)=(273.8^{\circ},-13.5^{\circ}$) i.e. only slightly inclined
from the plane used in this analysis. Our aim is not to reproduce
exactly the orbit of the Sgr galaxy, but to obtain systems close
enough so as to explore the luminosity profiles of the tails.

Particles are defined as bound, or unbound, depending on whether their
kinetic energy exceeds, or not, the satellite's internal potential
energy determined by {\tt SuperBox}. Leading and trailing arm
particles are then selected based on their orbital energies relative
to the orbit of the centre of mass. In the inner regions of the
satellite, approximately one half of the bound particles are assigned
to the leading arm and one half to the trailing arm. The profiles are
constructed in a fashion analogous to that used in
\citet{NiedersteOstholt:2010p2890} counting stars from a given arm in
angular bins covering $1.2^{\circ}$ on the sky. The bin size is chosen
to have sufficient resolution along the tails whilst avoiding
significant bin-to-bin variations. The counts are normalised by the
length along the stream that is covered by a given bin and corrected
for the angle that the stream makes with the radial direction at that
location.

In addition, we require that the one-component progenitor has lost
$\sim 60\%$ of its mass, as suggested in
\citet{NiedersteOstholt:2010p2890}, which allows us to identify the
appropriate pericentric passage (the sixth) to match to the
present-day data. The timescale over which the disruption is followed
is 4.1 Gyrs.  The two-component model is allowed to evolve through the
same number of pericentric passages, and loses $\sim 70 \%$ of its
mass along the way. As it evolves onto a tighter orbit, the time to
the sixth pericentric passage is slightly shorter at 3.6 Gyrs.

\begin{figure*}
	\centering
 	\subfigure[]{
	\includegraphics[width=0.4 \textwidth]{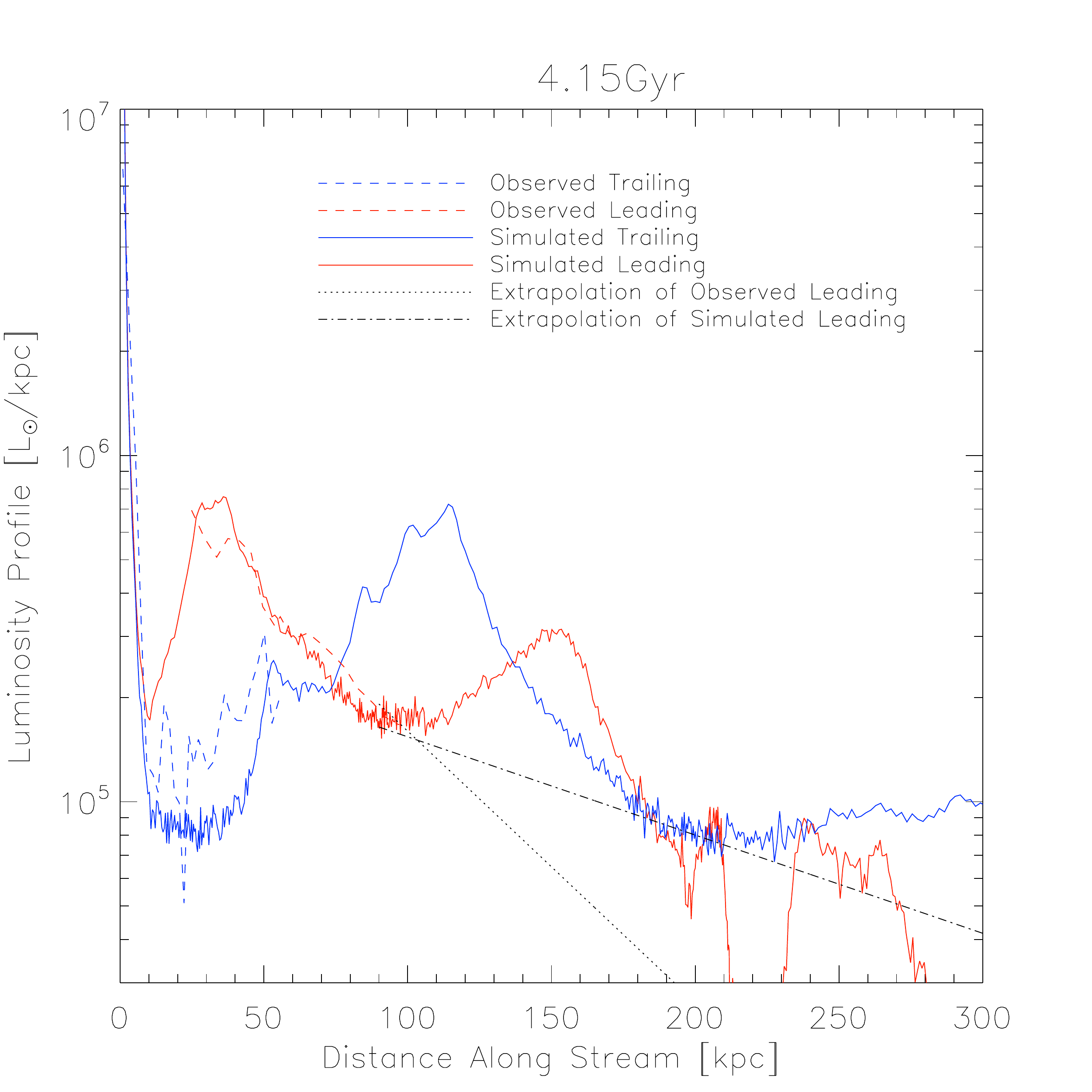}}
 	\subfigure[]{
	\includegraphics[width=0.4 \textwidth]{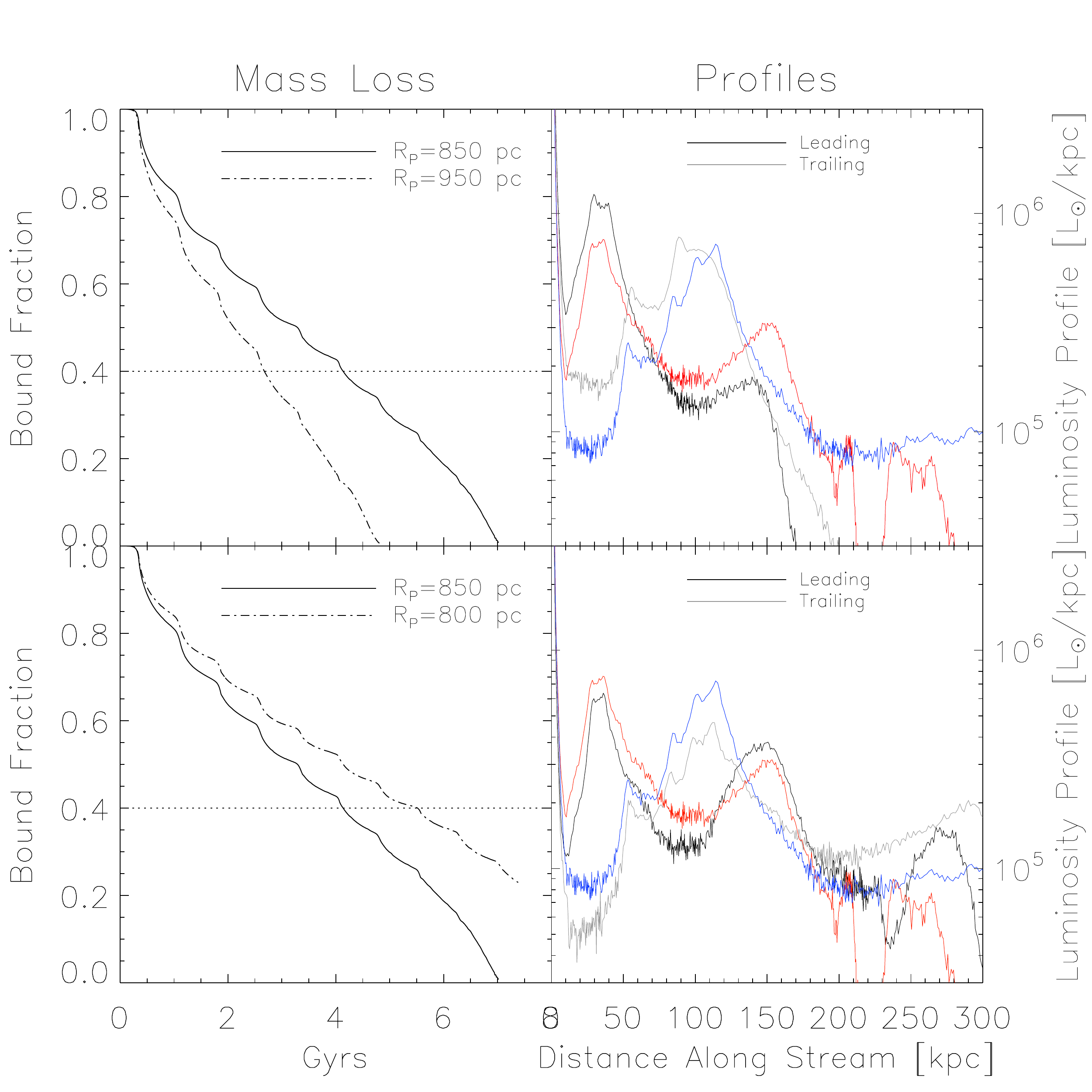}}
      \caption[Comparing Simulations to Observations]{(a): Leading
        (red) and trailing (blue) arm profiles after 4.15 Gyr for the
        satellite used by \citet{Law:2010p2807} with
        $(M,\RP)$=$6.4\times10^8$ $M_{\odot}$, $850$ pc. The snapshot
        is taken after passing pericentre and the satellite has lost
        $\sim60\%$ of its stars. The agreement between the simulated
        profiles (full lines) and observations (dashed lines) is good
        though the trailing arm appears slightly worse than the
        leading arm due to the log-scale). The short-dash dotted line
        shows the extrapolation used in
        \citet{NiedersteOstholt:2010p2890}. The long-dash dotted line
        shows an extrapolation based on the simulated profile, leaving
        out the additional bump in the profile.
        \citet{NiedersteOstholt:2010p2890} may have underestimated the
        amount of light not seen in the leading arm by as much as
        $~80\%$, though this does not change the luminosity range
        proposed for the progenitor Sgr since the extrapolation
        accounts for only a minor fraction of the total
        luminosity. (b): Varying the size of progenitor results in a
        poorer match with the observations, as changing the density of
        the progenitor influences how quickly stars disperse along the
        orbit. The left-hand column shows the mass-loss history of a
        satellite with $\RP$ = 950 pc (top) and 800 pc (bottom), both
        with mass $6.4\times10^8$ $M_{\odot}$, compared to the
        benchmark satellite. The right hand column shows the
        corresponding density profiles when the satellite has lost
        $60\%$ of its mass. The benchmark model is in red or blue
        (leading or trailing) whilst the larger and smaller
        scalelength models are in grey.  There are noticeable
        differences in the density distribution.}
	\label{fig:lawmodels}
\end{figure*}

\begin{figure*}
	\centering
	\subfigure[]{
	\includegraphics[width=0.45 \textwidth]{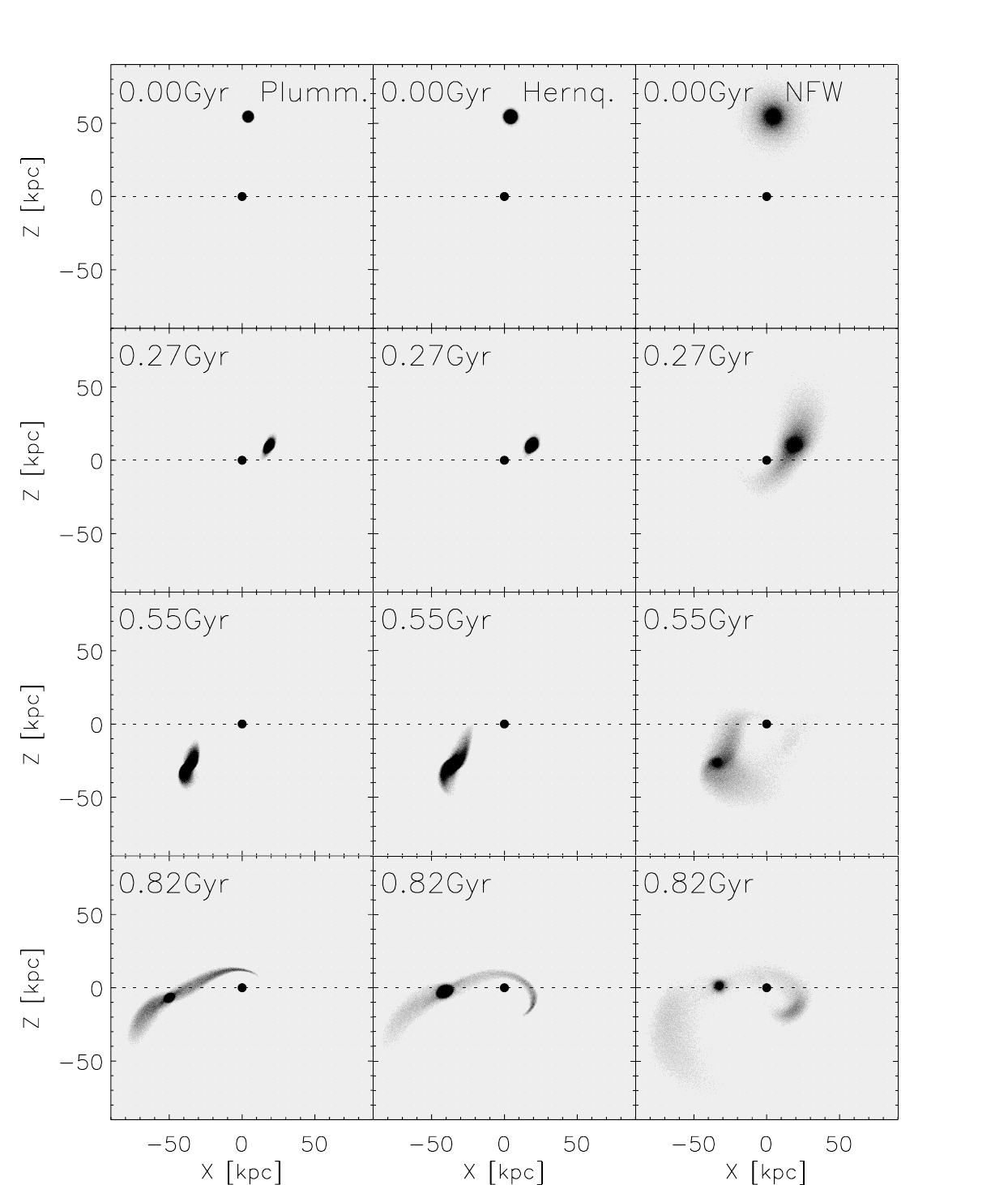} }
	\subfigure[]{
	\includegraphics[width=0.45 \textwidth]{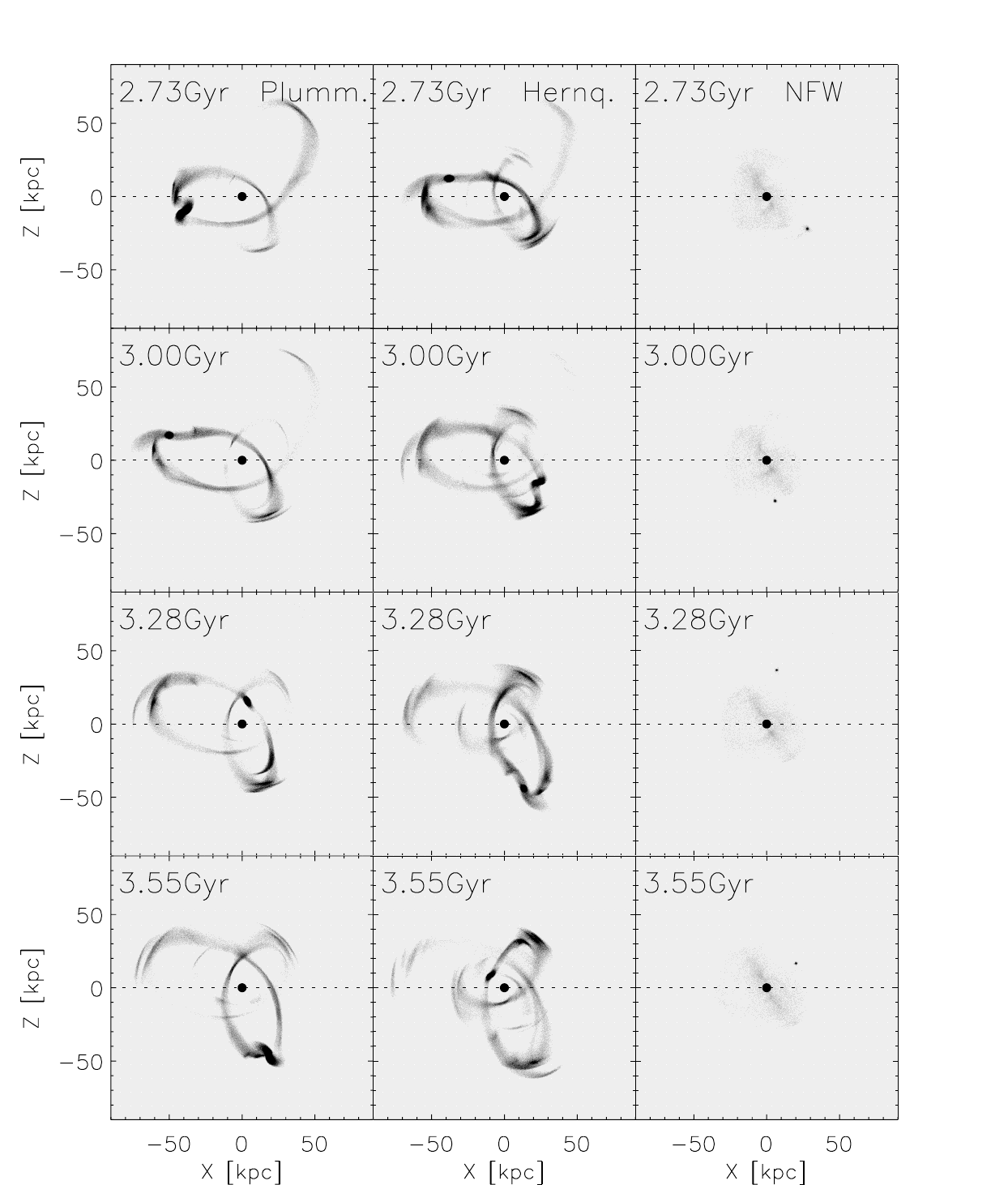} }
      \caption[]{A comparison between the disruption of the one and
        two-component models (a) for the first 0.8 Gyrs and (b) at
        late times. The plots run forward in time from left to right
        and top to bottom. In both panels, the left-hand column shows
        the one-component Plummer sphere disrupting in a Milky Way
        potential. The middle column shows the baryonic component of
        the two-component model, whilst the right-hand column shows
        the evolution of the dark matter component. The extended dark
        matter begins disrupting immediately. Compared to the
        one-component model, the baryons in the two-component model at
        first disrupt more slowly. This results in lower
        density, longer tails --- compare, for example, the baryonic
        material in the two component model with the one-component
        model at 082 Gyr. However, once a significant
        fraction of the protecting dark matter is removed, the
        baryonic disruption accelerates. }
	\label{fig:3densities}
\end{figure*}

\section{The First Telling}

\subsection{Luminosity Profiles of the Streams}

Figure \ref{fig:snaps} shows snap-shots from a typical disruption
of the one-component model, as well as the corresponding density
profile along the stream. During the initial phase of the disruption,
leading and trailing arms look very similar and their density profiles
lie very nearly on top of each other. As the disruption proceeds and
the progenitor and debris travel around the orbit, the leading and
trailing arms develop differently, having different lengths and
densities in a given snap shot.

The apocentric distances of the leading and trailing arm debris are
approximately $44$ kpc and $80$ kpc respectively.  These differences
can be attributed to the different speeds with which unbound particles
travel at various locations along the orbit \citep[see
e.g.][]{Dehnen:2004p7,Kupper:2010p1705} and the offset in energy
between the leading and trailing arms relative to the orbit of the
centre of mass. On a given orbit, particles close to pericentre travel
fastest resulting in lower densities and more stretched out tidal
streams. By contrast, particles spend more time near apocentre and the
tails appear compressed and have higher densities. This can be seen,
for example, in the snapshots at $3.51$ Gyr which are taken when the
main body of the progenitor has just passed its orbital pericentre (at
$3.32$ Gyr). The leading arm is approaching the next apocentre and is
compressed. The trailing arm particles on the other hand have just
left the previous apocentre and are speeding up as they move towards
pericentre, and thus have a lower density. This is consistent with the
observed shape of the Sgr luminosity density profile, where the
leading arm has a significantly higher density than the trailing arm
\citep[see Figure 10 of][]{NiedersteOstholt:2010p2890}.

A progenitor with a different mass or size will, on a given orbit,
disrupt at a different rate and the resulting debris will spread along
the orbit with a different speed, depending on its tidal radius at
pericentre, which controls the scale over which the satellite loses
mass \citep{Johnston:2001p2478}. At fixed $\RP$ (and in an identical
host Galactic potential), a more massive progenitor loses mass more
slowly, as it is more tightly bound. However, the unbound particles
disperse along the orbit quicker, allowing the tails to grow longer,
due to the larger spread in energies in the unbound particles. The
slower mass-loss and quicker dispersion result in a lower
density. 

If identical one-component satellites are placed on different orbits
with the same pericenter, then the debris spreads out over larger
distances before reaching orbital turning points if the apocentre is
larger. So, more eccentric orbits generally have a lower density,
since the debris spreads out over a longer orbital path before being
compressed at apocentre.

Figure \ref{fig:lawmodels}a compares the observed profiles (dashed
lines) to the simulated profiles produced by disrupting of the
\citet{Law:2010p2807} Plummer sphere. The number of particles is
converted to luminosities by assuming $M/L=3.2$. This implies that the
progenitor's total luminosity is $\approx 2\times10^8 L_{\odot}$,
which is at the higher end of the range determined by
\citet{NiedersteOstholt:2010p2890}. The behaviour of the observed
leading and trailing arms is well reproduced by the simulations.
Figure \ref{fig:lawmodels}b shows, progenitors with the same total
mass, but with different sizes (800 pc $<\RP <$ 950 pc), yield
profiles that are different from our best-fit model. Consequently,
there is significant disagreement with the observed profiles.

The leading and trailing arms have very nearly the same number of
particles throughout the simulations. According to
\citet{Choi:2007p2799}, the self-gravity of the disrupting satellite
can lead to asymmetries in the amount of material in the leading and
trailing arms. However, our Sgr models are at the low-mass end of the
satellites considered by \citet{Choi:2007p2799} and hence are unlikely
to display much asymmetry between the leading and trailing arms. We
conclude that calculating the total luminosity of Sgr progenitor based
on measuring the luminosity of one tail only \citep[as done
in][]{NiedersteOstholt:2010p2890} does not introduce substantial
error.

The observed asymmetry in density between the leading and trailing
arms can be attributed to the phase of the orbit at the time of
observation. Shortly after pericentre, part of the leading arm is
compressed as it approaches the next orbital apocentre. The trailing
arm on the other hand is stretched out as its particles are still
approaching pericentre. There is a pile-up in the trailing arm,
stemming from particles at the previous apocentre.

The time taken to populate the stream ($\sim 4$ Gyr) is in reasonable
agreement with other predictions of the age of the debris
\citep{Law:2005p522,Law:2010p2807}. This is not surprising, since the
satellite and potential are essentially the same as that of the Law et
al. simulations. The simulations of \citet{Lokas:2010p3684} find a
significantly shorter disruption time at $1.25$ Gyr. This difference
likely arises because they focus on reproducing the properties of the
core of the progenitor rather than the tidal streams. As the authors
highlight, the shorter interaction timescale is not necessarily in
conflict with the longer interaction times required to form the tidal
tails, but rather suggests that the length of time over which the Sgr
core has been tidally stirred is relatively short.

\subsection{The Assumptions of Niederste-Ostholt}

\citet{NiedersteOstholt:2010p2890} reassembled the stars in the Sgr
remnant and tidal streams to estimate the luminosity of the Sgr
progenitor. They found that the luminosity is in the range (9.6-13.2)
$\times 10^7 \L_\odot$ or $M_V ~ -15.1$ to -15.5, making the Sgr
progenitor comparable to, but somewhat less luminous than, the
present-day Small Magellanic Cloud.

However, \citet{NiedersteOstholt:2010p2890} made several assumptions,
particularly at locations where stream data was sparse or
unavailable. These assumptions can now be tested using the N-body
models. They include: (i) the behaviour of the leading arm near the
core where it is obscured by the Galactic disc; (ii) the extrapolation
of the leading arm beyond where it is observed in the SDSS data; and
(iii) the rising density of the trailing arm after an initial drop
resulting from the assumed metallicity gradient between the core and
tails. In this section, we comment on each of these assumptions based
on our simulations.

An important open question that remained in the observed profile of
Sgr is the behaviour of the leading arm in the region where it is
obscured by the Milky Way disk. This is the region at a distance along
the stream less than $\approx 20$ kpc in Figure \ref{fig:lawmodels}a
where the observations of the leading arm (red dashed line) end. From
these simulations, it seems much more likely that the leading arm
profile of Sgr dips down and joins the trailing arm profile near its
minimum, rather than a continuing exponential rise suggested by the
limited observations. This is borne out in other progenitor models
(175 pc $< \RP <$ 700 pc, $10^7< M <10^9$ $M_{\odot}$) and other
orbits with pericentre at 15 kpc but apocentric distances between
35-100 kpc. Based on the drop in density and the obscuring influence
of the Galactic disk, it will be challenging to detect this portion of
the leading arm.

In Figure \ref{fig:lawmodels}a, the leading arm follows an exponential
decay save for another bump in the profile occurring at the apocentre
an orbit ahead of the current location. If we look only at the
extrapolated portion of the tail (distance $>100$ kpc), then including
the bump increases the amount of light by $~70\%$ over the exponential
extrapolation. However, since the extrapolated portion of the profile
contains relatively little light, the total luminosity in the tail
changes only by $~5\%$ when integrating the extrapolated profile
versus integrating the actual profile, indicating that ignoring the
additional bump has only a small effect. The amount of light in the
extrapolated portion of the tail is much more sensitive to the slope
chosen. The exponential decline of the leading arm assumed in the
determination of the total luminosity of Sgr may have been too
steep. Employing a shallower extrapolation of the leading arm profile
increases the total amount of light in the extrapolated portion of the
leading arm by as much as $\sim 80\%$. Including a correction for the
bump at $\sim 150$ kpc along the stream adds another $\sim 25\%$.

Concerning the trailing arm, \citet{Majewski:2003p337} found their
density profile based on 2MASS M-Giants to be flat. By contrast,
\citet{NiedersteOstholt:2010p2890} found the trailing arm of the Sgr
stream has a positive slope due to the population differences between
the satellites core and tails reflected in the changing ratio of the
number of M giants to total luminosity. This population difference
indicates a possible metallicity or age gradient between different
regions of the undisrupted dwarf.  The existence of such a gradient is
already observed by \citet{Chou:2007p744} who find a mean metallicity
difference of $\sim0.6$ dex in [Fe/H] between the core and the
stream. The fact that the simulations reproduce this behavior suggests
that the proposed population gradient is in fact an acceptable
approximation. However, we also need to consider that the density of
the trailing arm is increasing when moving away from the core due to
the pile-up of particles at the previous apocentre in the orbit,
towards which the end of the trailing arm is reaching. The discovery
of a peak in density of the trailing arm followed by a drop would
support this view.

Considering the insights gained from the simulations concerning the
light in the leading arm farther than $100$ kpc along the stream as
well as the behaviour of the leading arm in the region obscured by the
Galactic disk, {\it we can update the luminosity range for the Sgr
  system from $9.6-13.2\times10^7 L_{\odot}$ to $9.9-14.4\times10^7
  L_{\odot}$}.

Within the uncertainties inherent in this approach, the luminosity
range for Sgr has changed only slightly and the overwhelming factor
controlling the spread remains the luminosity function assumed for Sgr
and its debris \citep[see][Figure 4.6]{NiedersteOstholt:2010p2890}.

\begin{figure}
	\centering
	\includegraphics[width=0.5 \textwidth]{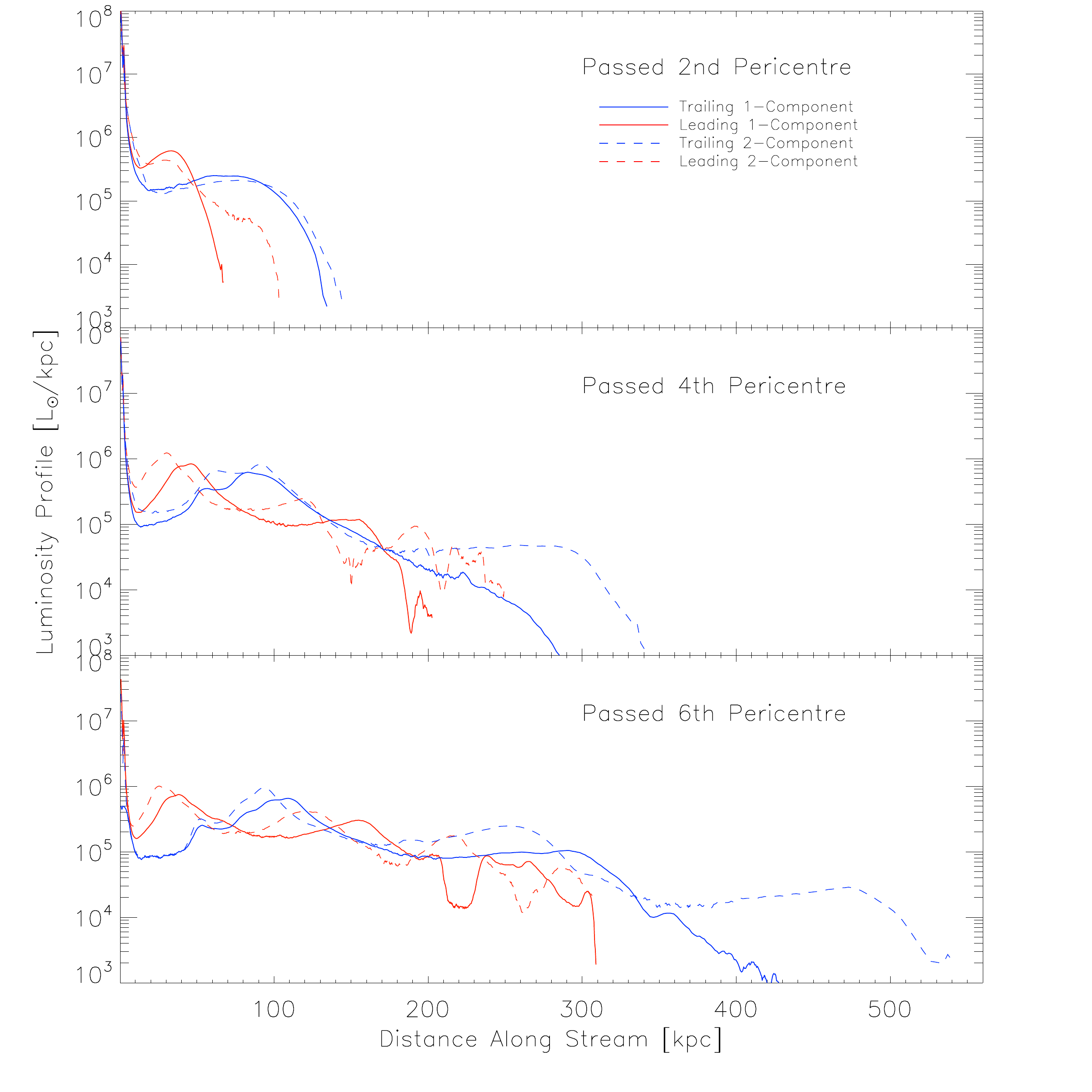}
        \caption[]{The luminosity profiles of the leading (red) and
          trailing (blue) tails for the one-component (solid lines)
          and two-component (dashed lines) models after (a) the second
          pericentric passage, (b) the fourth pericentric passage, and
          (c) the sixth pericentric passage. It is noticeable that at
          early times the two-component model is somewhat less dense
          than the one-component model, but has a much more extended
          central region and significantly longer tidal tails.  After
          four pericentric passages, the densities are already very
          similar, though the two-component model has longer
          tails. Finally, after six pericentric passages even the
          density in the central region of the two-component satellite
          matches that of the one-component model closely. Slight
          differences remain along the profile resulting from the
          debris spreading farther along the orbit.}
	\label{fig:modcomp}
\end{figure}

\section{The Second Telling}

Dwarf galaxies show ample evidence of being dark matter dominated. So
far our models have ignored the possibility that the stars visible
from the Sgr galaxy were initially embedded in a dark matter
halo. This motivates our second telling of the tale of the Sgr galaxy.

Using the code employed in Fellhauer et al. (2007), we generate a
two-component galaxy consisting of baryons distributed in a Hernquist
sphere embedded in an NFW dark matter halo. The two-component model is
released in the Milky Way potential with the same initial conditions
as the one-component model and luminosity profiles for the baryons are
produced in the same fashion.

Specifically, the stellar distribution is represented by a Hernquist
sphere of $10^6$ particles. It has a mass of $6.4\times10^8 M_{\sun}$
and a scale radius of 850 pc, thus closely matching the Plummer model
employed above. The NFW halo is represented by $5\times10^5$ particles
with a scale radius of $3.5$ kpc and a virial radius of $52.5$
kpc. This concentration ($c \approx 15$) results in the baryonic
component having a starting mean velocity dispersion of $\sim 19$
kms$^{-1}$. This is larger than the initial velocity dispersion of the
one-component model. As the disruption takes place,the mean velocity
dispersion declines to a value that lies within the observational
bounds at the present epoch.

Figure \ref{fig:3densities} shows the tidal evolution of the two model
satellites considered in this paper. The left-hand column shows the
one-component Plummer model for comparison purposes. The middle and
right-hand columns show the evolution of the two-component model's baryonic
and dark matter distributions respectively. Notice that the
dark-matter NFW halo begins to disrupt immediately. The presence of
dark matter increases the baryons' tidal radius. On the one hand, this
initially inhibits the disruption of the baryonic component, when
compared to the evolution of the Plummer model.  However, the
increased tidal radius translates into an increased energy spread in
the debris. This results in thicker tails and stars dispersing faster
along the orbit populating the tails out to longer distances. The dark
matter halo quickly becomes significantly disrupted (after ca. two
orbits) resulting in an accelerated rate of disruption of the baryonic
component. This effect is increased by a slight orbital decay
experienced by the two-component satellite once the dark-matter
disrupts (the pericentre shrinks from $\approx 15$ kpc to $\approx 11$
kpc).

The leading and trailing tail profile shapes at the snapshots taken
after then second, fourth, and sixth pericentric passage are shown in
Figure \ref{fig:modcomp}. The sixth passage corresponds to the one
chosen above as a comparison to the observations with Sagittarius.  In
this case, the final profiles of the two-component model look similar
to those of the one-component model, albeit longer. Since the
agreement between the profiles is good near the satellite core, the
two-component and one-component models are almost indistinguishable
based on the currently observed data.

Nonetheless, deeper data could distinguish between the profiles. The
baryonic material in the two-component model starts off with a higher
mean velocity dispersion than the one component-model ($\sim 19$
kms$^{-1}$ as opposed to $15$ kms$^{-1}$), though both remnants have
essentially the same mean velocity dispersion at the end. This however
means that the stars in the tails produced by the two-component model
are moving on average somewhat faster and so the tails grow longer
over the disruption timescale. At the present orbital phase, this is
especially the case for the trailing arm. One possible discriminant
between the models might be to search for this extended trailing tail
using giant stars or blue horizontal branch stars.

There are also differences in the properties of the remnant, which is
more compact and rounder in the two-component model than the
one-component, (though neither model is as flatted as the actual Sgr
remnant).  The ratio of initial to final baryonic mass within a 5 kpc
aperture is 0.43 for the one-component case and 0.29 for the two
component. This shows that more of the baryonic material is removed
from the immediate vicinity of the Sgr remnant in the latter case.

\section{Discussion and Conclusions}

We have used the particle mesh-code {\tt SuperBox} to disrupt models
of the Sagittarius satellite in a Milky Way potential. As progenitors,
we constructed N body representations with similar baryonic
components, but with very different dark matter content. The
one-component models mimic a situation in which any dark matter has
been stripped from the dwarf galaxy prior to the disruption of the
baryonic component, or in which the satellite had very little dark
matter in the beginning. The two-component models have a baryonic
component embedded within a cosmologically-inspired dark halo.  Their
disruption in the Milky Way potential provides two tellings of the
tale of the formation of the Sgr galaxy and its tidal streams.

The stellar components are structurally indistinguishable at outset,
having similar characteristic lengthscales and masses. They have
different initial mean velocity dispersions, but this ensures that the
remnant's final velocity dispersion lies within the present-day
observational constraints. We have provided the first comparisons
between simulation data and the observed luminosity density profiles
of the leading and trailing Sgr tails constructed by
~\citet{NiedersteOstholt:2010p2890}.

Our main conclusions are

\medskip
\noindent [1] The profiles of the tails can provide a strong
constraint on the dark matter content of the Sgr system, though this
requires deeper data than is currently available. The two-component
model has longer tails than the one-component model. This is because
the tidal radius and the starting mean velocity dispersion of the
baryonic component in the two-component model are larger than that of
the one-component model. Stars diffuse slightly more quickly along the
tails, which as a consequence are longer and less dense. The length of
the tails is an important discriminant. If we do observe very long
tails for Sgr, then a one component model would need to be more
massive to populate them to those distances. Even though there is some
freedom to vary the (stellar) mass to light ratio by a factor of $\sim
2$ from our assumed value of $3.2$, we have verified by simulations
that this does not permit the present data on the tail length and
profiles to be matched.

\medskip
\noindent [2] One-component models successfully reproduce the density
of the tails with only a baryonic component.  Two-component models
produce similar results within $180^\circ$ of the remnant. Although
the addition of a dark matter halo does initial protect the baryons
from disruption, once the dark matter halo has been removed, the
profiles develop in a similar fashion. With the present data , it is
not possible to distinguish between one and two-component models based
on fitting the profiles.

\medskip
\noindent [3] The simulations support the idea that the amount of
light in the leading and trailing arms is nearly identical. The main
reason for the difference in luminosity density between the leading
and trailing arms observed in Sgr is the the orbital phase at the
present epoch.  Shortly after pericentre, part of the leading arm is
compressed as it approaches the next orbital apocentre. The trailing
arm on the other hand is stretched out, as its particles are still
approaching pericentre.

\medskip
\noindent [4] The simulations have strengthened a number of the
assumptions made by \citet{NiedersteOstholt:2010p2890} in determining
the total luminosity of the Sgr system. It is very likely that the
leading arm's luminosity density, in the region where it is obscured
by the Galactic disk, dips down to become symmetric with the trailing
arm. Furthermore, the extrapolation of the leading arm's profile
proposed in \citet{NiedersteOstholt:2010p2890} is appropriate (with
the possibility of a slightly different slope for the exponential
decline), but requires the inclusion of additional light due to
pile-ups at the next orbital apocentre. These factors lead to a slight
upward revision of the total luminosity of the Sgr progenitor proposed
by \citet{NiedersteOstholt:2010p2890} to $9.9-14.4\times10^7
L_{\odot}$.

\medskip
\noindent [5] The pile-ups of particles in the leading arm seems to be
as dense as some parts of the already observed profile. Hence, it may
be possible to detect these in existing survey data. Similarly, the
increasing density of the trailing arm may allow further detections of
that debris tail. The number of such density enhancements discovered
may provide valuable constraints on the length of the tidal tails and
the number of orbits over which the debris is spread. However, the
highest density enhancements occur naturally around the apocentre of
the orbits, which may be as far as $90$ kpc from the observer in the
case of the trailing arm (e.g., Koposov et al. 2011). This may render
detection difficult with current surveys such as the SDSS. The highest
overdensity in the leading arm, at a distance of $\sim 47$ kpc, has a
main-sequence turn-off at $i\sim21.4$. If we assume that turn-offs can
be detected down to $i=22$, this suggests that we may be able to
detect such overdensities out to $\sim 60$ kpc using this
approach. Other stream tracers, such as blue horizontal branch stars
or M giants, may be observed out to greater distances.

\section*{Acknowledgments} 
We thank the referee, David Law, for a careful reading of the paper
and valuable comments. MNO thanks the Gates Cambridge Trust, the Isaac
Newton Studentship fund and the Science and Technology Facilities
Council (STFC), whilst VB acknowledges financial support from the
Royal Society.

\bibliography{citations_sgr_sim}

\label{lastpage}

\end{document}